\documentclass[12pt]{article}
\pdfoutput=1
\usepackage[utf8]{inputenc}
\usepackage[english]{babel}
\usepackage{extarrows}
\usepackage{amsmath}
\usepackage{amsfonts}
\usepackage{amssymb}
\usepackage{graphicx}
\usepackage{fourier}

\usepackage{caption}
\usepackage{subcaption}
\usepackage{float}
\usepackage{color}
\usepackage{abstract}
\usepackage{appendix}
\usepackage{authblk}
\usepackage{xcolor}

\numberwithin{equation}{section}

\usepackage[hidelinks]{hyperref}

\newcommand\be{\begin{equation}}
\newcommand\ee{\end{equation}}
\newcommand\ba{\begin{eqnarray}}    
\newcommand\ea{\end{eqnarray}}      

\title{Signature change as phase transition in holography}

\author{Marcelo Botta Cantcheff \footnote{botta@fisica.unlp.edu.ar}}

\affil{Instituto de F\'isica La Plata - CONICET and 

Departamento de F\'isica, Universidad Nacional de La Plata 

C.C. 67, 1900, La Plata, Argentina}

\date{March 30, 2025}
\usepackage[left=2cm,right=2cm,top=2cm,bottom=2cm]{geometry}

\begin{document}

\begin{titlepage}

\maketitle

\hspace{1cm}\emph{\small{Essay written for the Gravity Research Foundation 2025 Awards for Essays on Gravitation}}

\begin{abstract}

In holographic quantum gravity, Euclidean pieces of the spacetime appear in the large $N$ limit as representing semi-classical states of the theory.
In this essay, we argue that the duals of entangled states are spacetime geometries that contain Euclidean regions in order to preserve classical connectivity. 
 Thereby, the proposal is to extend the ER-EPR conjecture to regimes whether wormholes (Einstein-Rosen bridges) become unstable but the entangled structure of the dual state persists.

\end{abstract}
\thispagestyle{empty}
\end{titlepage}

\section{Introduction}

Underlying Hartle and Hawking's original proposal is the idea that spacetime can undergo radical changes in its structure near singularities like the Big Bang, such that its effective signature becomes Euclidean rather than Lorentzian, leading to the robust \emph{no-boundary} proposal \cite{HH}. In this essay, we aim to explore this possibility from a holographic perspective and emphasize that, in certain regimes, the holographic spacetime can effectively include Euclidean regions.

The AdS/CFT correspondence represents the paradigmatic realization of holographic duality where spacetime with fixed (AdS) asymptotics can be viewed
as emergent from a ordinary quantum field theory defined on its conformal boundary (CFT)\cite{adscft}.
In this framework the CFT is formulated as an $SU(N)$ gauge theory, providing a nonperturbative description of quantum gravity (QG), where the effective 
Newton constant $G_N$ is proportional to $N^{-2}$. Consequently, a generic CFT state encodes the information about an asymptotically AdS (aAdS) spacetime geometry, whose classical properties are recovered in the large-$N$ limit.

Since the remarkable observation \cite{vanram}, is commonly accepted that \emph{entangled} states in the CFT should be dual to some type of classically connected asymptotically AdS (aAdS) spacetimes, and the connectivity is precisely due to the entanglement property of the states. 
Other aspects of CFT states, including their entanglement features, and the rules linking them to dual geometry properties such as topology, causal structure, and signature, remain unclear and warrant deeper investigation \footnote{Recent advances have been made on the emergent spatial topology and coherence properties of states \cite{botta22}}. This work focuses on the emergent signature

In what follows we analyze the Hawking-Page phase transition scenario, to show that in holographic (quantum) gravity Euclidean regions of the spacetime must be effectively taken into account. In particular there are entangled states (EPR) in the dual field theory, whose dual spacetime necessarily contains connected regions of Euclidean signature. A key aspect of this argument is that such Euclidean regions would appear when the conditions required for a non-traversable Lorentzian wormhole (an Einstein-Rosen bridge) cannot be fulfilled. Therefore, in the context of the ER-EPR proposal, it suggests that (ER) Lorentzian/Euclidean can be understood as two  \textit{phases} of the dual geometry.

\section{Entanglement and classical connectivity}

 For simplicity, we will consider scalar CFT operators ${\cal O}$, dual to a massive scalar field $\Phi(x)$ of mass $m$ defined 
 in asymptotically AdS$_{d+1}$ spacetime. The conformal weight of operator, $\Delta$, is related to the mass of
 the bulk field by $m^2=\Delta (\Delta-d)$.  In the large $N$ limit the propagators between two insertions on the bulk are given by the path 
  integral in the bulk geometry $M$ \cite{correl1,correl2}:

\be \label{correl} \langle \Psi | {\cal O}(x) {\cal O}(y)| \Psi \rangle =  \int_{\gamma  \subset  M} [D\gamma] \; e^{im\, l[\gamma(x,y)]} \;\,.\ee
This stands for a sum over all the continuous curves  
 in a (in principle) Lorentzian spacetime manifold $M$ with endpoints $x,y$ on the boundary. The functional $l[\gamma(x,y)]$ is the proper length of the curve and for spatially separated points, it is purely imaginary.
For high conformal weight $m\sim \Delta$, the left hand side can be approximated by \be\label{correlwkb} \sum_{k} e^{i \Delta\, l[\gamma_k(x,y)]} \qquad,\ee
where the sum reduces only to the contribution of the geodesic curves $\gamma_k$.
They can be viewed as particle paths between the two boundary points.

Notice that the classical connectivity is defined by the existence of continuous curves in a spacetime manifold $M$, so this expression describes the statement \cite{vanram} manifestly. 
In fact, 
if $x$ and $y$ are points causally disconnected in the quantum field theory defined on the boundary manifold,
they do not interact; thus, l.h.s.(1) $ \neq 0$ implies that these points are connected by curves through the bulk, which cannot be timelike everywhere.

Conversely, existence of curves with pieces $-\infty< il < 0$ implies nonzero correlators on the left hand side, which can only be due to the entangled structure of $\Psi$ \cite{vanram}.
It is worth emphasizing that these arguments are crucial for our discussion.

\section{ER-EPR in holography}

The conjecture ER-EPR can be considered a further step in this sense \cite{er-epr}. This relates Einstein-Rosen (ER) non-traversable wormholes to highly entangled states states of certain quantum system, which in holography can be elegantly explained as follows \cite{sonner14}. 

Let us consider two non-interacting identical copies 
of  CFT (labeled by a subindex $1,2$) on $B_{1,2} \equiv S^{d-1} \times {\mathbb R}$, the Hilbert space is ${\cal H}_1\otimes{\cal H}_2$ and depending on the state, their gravity dual shall be asymptotically $AdS_{d+1}$ spacetimes. Let $x,y$ two arbitrary causally-disconnected points on $B_1$ and $B_2$ respectively.
 
If the state $\Psi\in{\cal H}_1\otimes{\cal H}_2$ is entangled (EPR), then the l.h.s. of eq. \eqref{correl} is nonzero, and the r.h.s of this equation implies 
the existence of curves in $M$ connecting $x$ and $y$, which cannot be time-like everywhere by virtue of our assumption of causal independence. Therefore, there are only the following two possibilities:

\begin{itemize}
    \item Lorentzian ER bridge: if the spacetime $M$ is assumed to be Lorentzian everywhere, it must have horizons that split the two asymptotic regions as in the Penrose diagram of Fig 1a, where the spatial slices (green line) are nothing but non-traversable ER wormholes.
    \item Euclidean region: the curves in eq.\eqref{correl}, $\gamma$ must traverse some region of $M$ with Euclidean signature.
\end{itemize}

We will see below how this possibility generalizes the ER-EPR statement to regimes where the EPR state cannot be dual to a Black Hole.

Notice that this argument implicitly assumes that the state, prepared via initial/boundary conditions in the bulk, satisfies conditions of energy and density consistent with the formation of a black hole, or a pair of entangled ones \cite{er-epr}.  
This remark is crucial for our argument below since, whether conditions are insufficient to give place to black holes, we should have regions of the space-time with Euclidean signature, in order preserving the connectivity of the dual geometry \cite{vanram}.

\begin{figure}[t]\centering
\includegraphics[width=.9\linewidth] {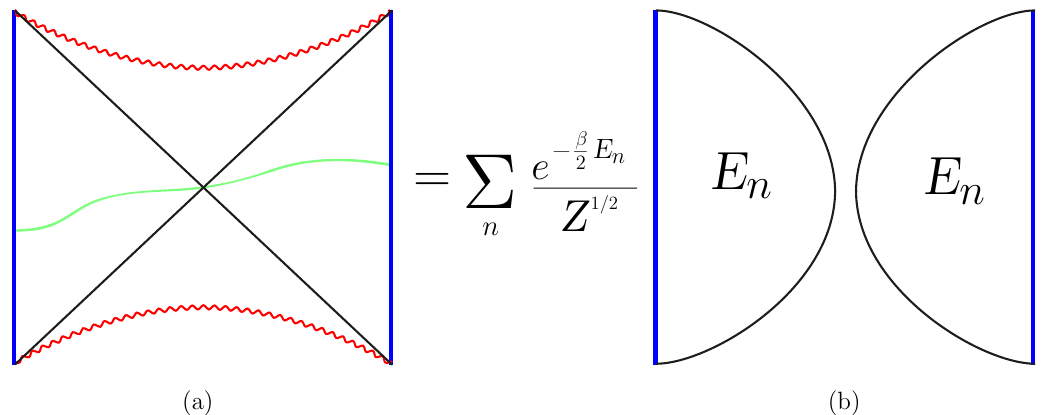}
\caption{\small{(a) Penrose diagram of a maximally extended AdS-black hole. The green line is a connected spacial slice representing the entaglement between the CFT's. (b) It represents the interpretation of \cite{vanram}: the linear superposition of states $|E_n\rangle_1 \otimes |E_n\rangle_2$ (dual to disconnected aAdS geometries), supposedly gives a connected spacetime. The blue lines represent the non-interacting CFT theories on the two asymptotic boundaries.} }
\label{vanram}
\end{figure}

\vspace{.5cm}
\section{Phase transition: ER vs Euclidean wormholes}
\vspace{.5cm}

Now consider the two (non-interacting) 
CFT identical copies living on  $S^{1} \times {\mathbb R}$, in the entangled (EPR) state:

\be\left|\Psi_\beta\right\rangle = \sum_n \, \frac{e^{-\frac{\beta}{2} \, E_n}}{Z^{1/2}}\left|E_n\right\rangle_1 \otimes \left|E_n\right\rangle_2
~\in{\cal H}_1\otimes{\cal H}_2~~,~~\beta\equiv(k_B T)^{-1}\label{BHstate}, \ee
where the $\left|E_n\right\rangle$ are a complete basis of eigenstates of the CFT Hamiltonian $H$, and $E_n$ are its eigenvalues.
 This is the known TFD state and from one of the two boundaries describes a thermal state of the quantum system at temperature $T$ \cite{tu,ume1,ume2}.
 Recall also that the state \eqref{BHstate} is invariant under the action of the combination $ H_1 - H_2$, although physical time evolution of the two systems is though to be generated by $ H_1 + H_2$ \cite{er-epr, eternal}.

\begin{figure}[t]\centering
\includegraphics[width=.9\linewidth] {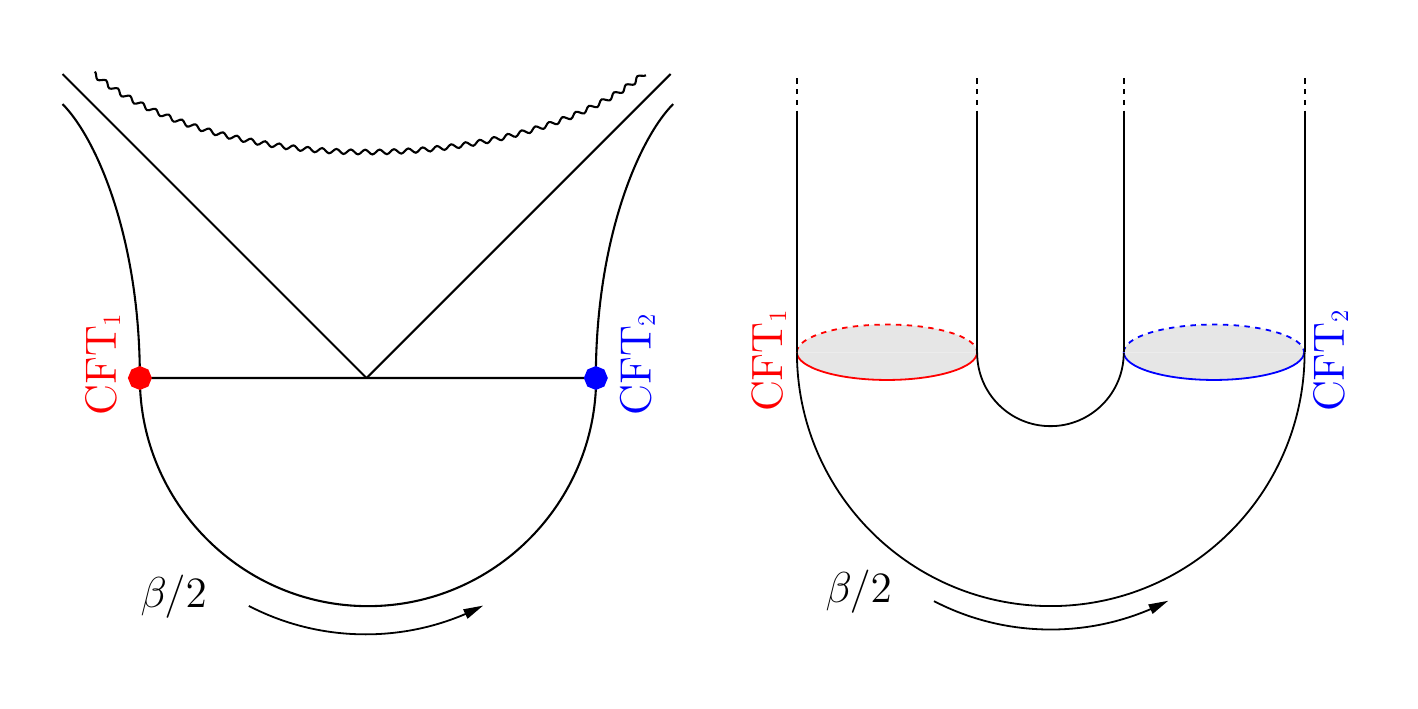}
\caption{\small{This figure represents the two geometries described by the state \eqref{BHstate}. The lower parts correspond to the (Euclidean) saddles of the Hartle-Hawking wave functional. In the figure on the left (a) the two-sided black hole is represented, the asymptotic regions are connected through the ER wormhole; on the right figure (b), the connectedness is realized through the Euclidean region} }
\label{HP}
\end{figure}

For sufficiently high temperature $\beta < 2\pi R_{AdS}$ the dual spacetime is the geometry represented in fig \ref{HP}(a), the ER-wormhole in according to the argument above. Nevertheless, if the temperature is low, such that the black holes are unstable \cite{haw-page, haw-page-witten}, the argument above
cannot work. As did before, let us consider that $x,y$ are points in the boundaries 1 and 2 respectively, 
then the explicit computation of the l.h.s. of eq. \eqref{correl} in the state \eqref{BHstate} is non-zero \cite{haw-page-witten, eternal, SvRL}, and thus, in order to satisfy that equation on the r.h.s., one must accept that there are curves $\gamma$ connecting these points through some Euclidean interval.
The dual space-time in this case can be represented as Fig1b, such that the spacetime connectivity is restored \cite{vanram}.

This proves our claim. The contrast with the previous argument that leads to ER-EPR, precisely lies in equation \eqref{correl}, because (except in most of the simplest cases) one cannot assume that $M$ is Lorentzian everywhere. In fact, as we consider non-trivial entangled states (ER) $\Psi$, on the gravity side we must take into account the Hartle-Hawking wave functional, defined as the gravitational path integral with Dirichlett boundary conditions on $\Sigma_\pm$ and the conformal boundary. Therefore, the geometry $M$ of \eqref{correl} must be promoted to $ M \equiv M_- \cup M_L \cup M_+$, which consists of three smoothly glued parts through the common surfaces $\Sigma_\pm$, $M_L$ is Lorentzian, and $M_{-/+}$ stands for the (Euclidean) saddle geometry and its respective time reflected (more details on this type of construction can be found in refs.\cite{SvRC, SvRL, us1, us3})

For instance, in the configuration that the real-time intervals shrink such that $M_L$ disappears, $M \equiv M_-  \cup M_+$ is the exact Euclidean AdS with time period $\beta$: the gravity dual of $\left\langle \Psi_\beta|\Psi_\beta\right\rangle$, and therefore, the (thermal) correlators can be computed following the conventional recipes \cite{haw-page-witten}.

Along this line one can compute precisely the correlators \eqref{correl} for the geometry of fig \ref{HP}(b).
by splitting the euclidean circle in two halves, and (smoothly) inserting two real time intervals parameterized by $t_1 $ and $ t_2$ \cite{eternal, SvRL, us3}. Thus, by solving for the field $\Phi$ in the (real-time extended) bulk partition function as usual, the detailed computation gives\footnote{ In our units $ R_{AdS}\equiv 1$.}:
\be
\nonumber\\
\langle\Psi_\beta|{\cal O} ( t_1 , \varphi_1) {\cal O}(t_2, \varphi_2) |\Psi_\beta\rangle =
\sum_{k\in\mathbb{Z}}\frac{(\Delta-1)^2}{2^{\Delta-1}\pi}\left[\cos\left((t_2 -t_1) + i\beta (k - 1/2)\right) -\cos(\varphi_2 - \varphi_1) \right]^{-\Delta}
\ee
This expression can be understand from the contribution of the possible paths connecting the points $x,y$ (r.h.s. of (1)), and according to \eqref{correlwkb}, the main contribution comes from the \emph{geodesic} paths that probe the Euclidean regions. The sum accounts for the winding number $k$ of these paths, although the shorter path dominates exponentially.

Note that if the $M_L$ shrinks completely ($t_1=t_2=0$), this agrees with the result previously known \cite{haw-page-witten}.

\vspace{.5cm}
\textbf{Discussion:}
 One can interpret the above transition scenario by imagining an ''experiment'' in which, if the temperature is lowered adiabatically on the ER bridge, such that the black hole becomes unstable, the two sets of degrees of freedom (1,2) inside AdS remain entangled, but geometrically connected to through an Eclidean region of space-time. This is literally represented in fig \ref{HP}(a) and \ref{HP}(b)

Although the existence of both euclidean saddles of the Hartle-Hawking wave functional, associated to the fundamental state (TFD) (or more general in/out) is known, the non-trivial observation here was that physical connection between the quantum ds.o.f. 1 and 2 can only be realized by curves propagating through the such Euclidean regions. The classical connectivity of space-time in entangled states as TFD (as $\beta \to \infty$), as demanded by \cite{vanram}, is not violated by virtue of these regions.

\section{Conclusions}

Curves connecting boundary points, as expressed in Eq. \eqref{correl}, are what actually probe the holographic spacetime, providing a useful notion of the effective emergent geometry. Based on this, we have argued that Euclidean spacetime regions can effectively emerge within holography.

Our reasoning is based on the idea that highly entangled states correspond to classically connected geometries. When macroscopic energy or temperature is sufficiently low to prevent black hole formation, spacetime connectivity \cite{vanram} is maintained. However, the cost is accepting the emergence of a region of spacetime with a Euclidean signature.

Although we have used the Hawking-Page context as laboratory, it is hoped that these features can be generalized, possibly to more general states of the quantum system where the energy and temperature are sufficiently low, or where the appropriate conditions for tunneling are met, leading to the emergence of a instanton-like dual geometry.

\vspace{.8cm}

 \textbf{Acknowledgements} The author is grateful to CONICET for financial support.

\end{document}